# Million-Q Dual-Polarization Micro-Fabry–Pérot Resonators in Silicon Nitride Photonic Integrated Circuits

Ziyang Xiong[a,1], Tong Lin[a,d,1,*], Liu Li[b], Hao Deng[a], Shihua Chen[c], Junpeng Lu[a,c], and Zhenhua Ni[a,c]

[a] *School of Electronic Science and Engineering, Southeast University, Nanjing, 210096, China*
[b] *School of Material Science & Engineering, Southeast University, Nanjing, 210096, China*
[c] *School of Physics, Southeast Univcersity, Nanjing, 210096, China*
[d] *Key Lab of Modern Optical Technologies of Education Ministry of China, Soochow University, Suzhou, 215006, China*
[1] *Ziyang Xiong and Tong Lin contributed equally to this work.*
[*] *lintong@seu.edu.cn*

A B S T R A C T

Miniaturized Fabry–Pérot standing-wave resonators and whispering-gallery travelling wave resonators constitute foundational building blocks for photonic integrated circuits. While both architectures offer transformative potential through high quality factors and dual-polarization operation, integrated Fabry–Pérot resonators face significant challenges in simultaneously achieving ultra-high Q-factors and broadband thermal tunability for fundamental transverse magnetic ($TM_0$) and transverse electric ($TE_0$) modes within a compact footprint—primarily due to polarization-dependent losses in conventional chip-scale reflectors. Here, we overcome this limitation by demonstrating an integrated silicon nitride dual-polarization micro-Fabry–Pérot resonator with polarization-insensitive Sagnac loop reflectors and multimode waveguides to effectively suppress losses and enable high-performances for both fundamental transverse magnetic ($TM_0$) and transverse electric ($TE_0$) modes. The device achieves record loaded quality factors of $2.38 \times 10^6$ ($TM_0$) and $3.48 \times 10^5$ ($TE_0$) respectively and intrinsic quality factors will be even higher. Moreover, both two modes are tuned over the whole free spectral range of around 0.111 nm ($TM_0$) and 0.112 nm ($TE_0$) with the thermal tuning efficiencies of approximately 1.04 pm/mW ($TM_0$) and 1.24 pm/mW ($TE_0$). These advances establish a new benchmark for compact, high-performance dual-polarization resonators in optical sensors, nonlinear and integrated quantum photonics.

## 1. INTRODUCTION

Integrated Micro-Fabry–Pérot (μFP) resonators made of highly reflective on-chip mirrors [1] achieve strong light confinement in a compact footprint while maintaining high quality-factors (Q) and narrow resonance linewidths. The resulting enhanced light-matter interactions and direct integration capability with photonic integrated circuits (PICs) enable their deployment in diverse applications, including semiconductor lasers [2], [3], [4], optical modulators [5], microwave photonics [6], and chip-scale frequency comb generation [7]. As the key performance metric, the Q-factor of a μFP resonator is fundamentally governed by its dominant loss mechanisms, primarily limited by reflection losses [8] and propagation losses [9]. Therefore, suppressing these losses is critical for achieving high-performance lab-on-a-chip systems [10], as evidenced by recent demonstrations of in-vacuum-bonded Fabry–Pérot cavities with Q-factors exceeding $4.26 \times 10^9$ [4].

The most challenging obstacle to the integrated ultra-high Q-factor μFP resonators is the near-unity reflectivity provided by the on-chip mirrors [1]. As the most prevalent reflector configuration, distributed Bragg reflectors (DBRs) rely on constructive interference in multiple sections to achieve high reflectivity. They represent the standard implementation for μFP cavities [11], [12], though their typical Q-factor remain constrained to the $10^4$ range due to interfacial scattering losses. To improve Q factors, Xu et al. propose geometrically engineered reflectors that substantially reduce reflection losses, pushing the Q factor beyond $2 \times 10^6$ with optimized 1.5-mm-long tapers [8]. Notably, current research on integrated μFP resonators predominantly targets TE modes, with only a single computational study addressing TM modes [13]. Specifically, high-Q TM-polarized μFP resonators exhibit distinct advantages for applications, such as optical sensors where the TM mode has a larger modal overlap with the analytes [14] and quantum cascade lasers of which intersubband transitions are inherently TM-polarized [15], [16]. The differential thermo-optic response of TE/TM dual-polarization modes can be used to compensate for laser frequency drift induced by temperature fluctuations, thereby significantly enhancing the accuracy and stability of frequency references—particularly in high-precision metrology and ultra-stable communication applications [17].Furthermore, the synergic TE-TM polarization behaviors empower simultaneous sensing [18], [19], [20] and multimodal perception [21]. The untapped potential of dual-polarization μFP resonators thus represents a critical frontier for PICs.

In this study, we present the design and experimental demonstration of an ultra-high-Q-factor dual-polarization μFP resonator based on polarization-insensitive Sagnac loop reflectors (SLRs), monolithically integrated on a $Si_3N_4$ photonic platform. This platform has become critical for PICs, owing to no two-photon absorption above 800 nm and propagation losses as low as 1 dB/m [22], [23], [24], [25], [26], [27], enabling integrated resonators with record-high Q factors. By precisely engineering the $Si_3N_4$ waveguide cross-section close to a square geometry, the μFP cavity simultaneously supports high-Q resonances in both TM and TE modes. The resonator incorporates two identical SLR-based mirrors, each constructed with a 3-dB directional coupler and a low-birefringence waveguide loop, significantly enhancing the reflectivity of the μFP cavity. Moreover, the adoption of an optimized multimode waveguide design effectively mitigates intracavity propagation losses. Experimental results show loaded Q-factors beyond $2.38 \times 10^6$ ($TM_0$) and $3.48 \times 10^5$ ($TE_0$) and the intrinsic Q-factors will be even higher. Under thermo-optic tuning, the device exhibits broadband spectral tuning exceeding one free spectral range (FSR), with tuning efficiencies of approximately 1.04 pm/mW ($TM_0$) and 1.24 pm/mW ($TE_0$). Such dual-polarization, broadly tunable, ultra-high-Q resonators hold substantial promise for advanced nonlinear photonics and integrated photonic applications.

## 2. DESIGN AND PRINCIPLE

We design the ultra-high-Q dual polarization μFP resonator based on an 0.8-μm thick $Si_3N_4$ photonic platform. The device is fabricated from the commercial $Si_3N_4$ foundry (LIGENTEC MPW-AN800-35), incorporating input and output waveguides of 1 μm width, capable of supporting fundamental TM and TE modes in the optical S-C-L-U-band. Figs. 1(a) and 1(b) depict the schematic and corresponding optical microscope images of the proposed device, in which the μFP cavity is composed of two SLRs interconnected by 40-μm-long tapered transition waveguides and a 3732-μm-long multimode waveguide. Notably, by employing the Bézier bend-assisted spiral layout, we can reduce the straight-segment length to approximately 200 μm, thereby significantly minimizing the footprint and enabling more flexible waveguide routing. The cross-sectional view of the multimode waveguide region within the μFP cavity is shown in Fig.1 (c), comprising a 3-μm-wide $Si_3N_4$ core on a 4-μm-thick silicon dioxide bottom cladding layer. Such a wide multimode design significantly reduces scattering losses induced by sidewall roughness, thereby substantially enhancing the intrinsic Q-factor and overall resonator performance. To suppress modal crosstalk in the multimode region, we introduce two 40 μm linear adiabatic tapers (width 1→3 μm) between the single-mode and multimode sections, enabling a low-loss $TE_0$ mode transition without exciting higher-order modes and ensuring stable single-mode propagation within the multimode region. Precise wavelength tunability is achieved through microheaters positioned approximately 1.7 μm above the $Si_3N_4$ layer, exploiting the thermo-optic effect to finely modulate the effective index and consequently adjust the resonant wavelength with high accuracy.

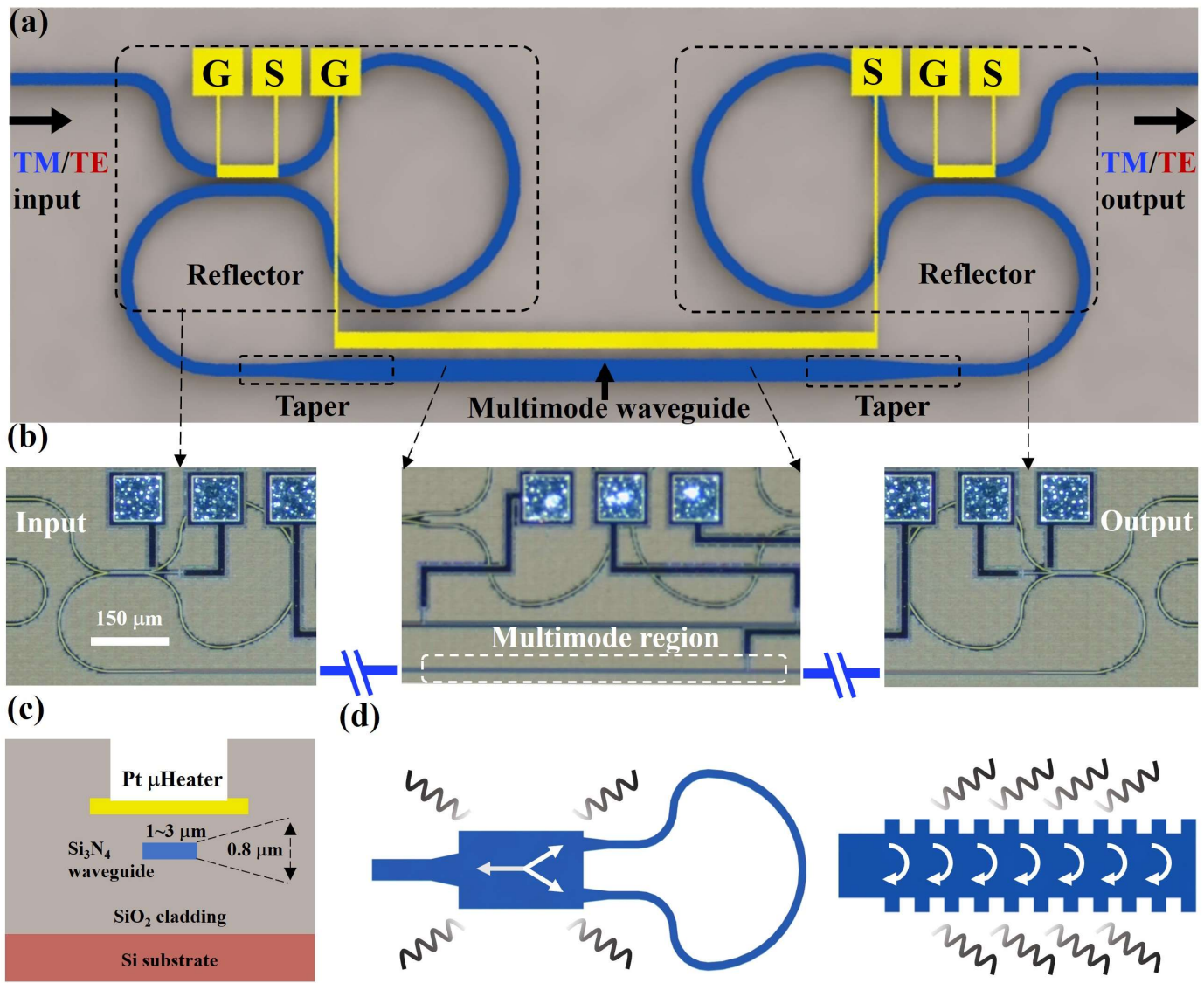

**Fig. 1** (a) Schematic view of the proposed ultra-high-Q dual-polarized μFP resonator with SLR-based mirrors on a silicon nitride platform. (b) Optical Microphotographs of the μFP resonator's three different sections: the left SLR, the middle multimode waveguide, and the right SLR; the scale bar is 150 μm corresponding to the electrical pad's horizontal spacing. (c) Cross-sectional view of the waveguide region of the μFP resonator. (d) Illustrations of the reflection loss mechanisms in one MMI-based Sagnac loop reflector and one DBRs, respectively.

We reduce the reflection losses of $TM_0$ and $TE_0$ modes by using polarization-insensitive SLRs. Commonly reported mirror structures based on multimode interference (MMI) and DBR inherently suffer from considerable scattering losses [28] and the grating-cavity coupling loss [29], as schematically illustrated in Fig.1(d), imposing fundamental constraints on device performance. In contrast, the SLR design based on a low loss 3-dB directional coupler [30] utilized here effectively mitigates scattering losses. Furthermore, the near square waveguide cross-section (1×0.8 $\mu m^2$) reduces the polarization birefringence, thus represents a more advanced and efficient approach for high-performance dual-polarized μFP resonator implementations.

The SLR is constructed using a 2×2 3-dB directional coupler whose two outputs are interconnected by two S-bends (75 μm radius) and one circular bend (100 μm radius), forming a closed-loop configuration. Consequently, the transmission and reflection coefficients of an individual SLR are given by:

$$t_{SLR} = (t^2 - k^2) a_1 e^{-j\beta_1 l_1} \quad (1)$$

$$r_{SLR} = -2jkta_1 e^{-j\beta_1 l_1} \quad (2)$$

where t and k denote the transmission and coupling coefficients of the 3-dB directional coupler within the SLR respectively, $a_1$ and $l_1$ represent the loss factor and physical length of the SLR, and $\beta$ is the propagation constant of the $Si_3N_4$ waveguide. When this directional coupler is ideal ($t=k=0.5$), 100% reflection can be achieved. The transmission function of the µFP cavity $T_{FP}(\lambda)$ can be expressed as follows:

$$T_{\mathrm{FP}} = \frac{t_{SLR}^2 a_2 e^{-j\beta_2 l_2}}{1 - r_{SLR}^2 a_2^2 e^{-j2\beta_2 l_2}} \tag{3}$$

where $t_{SLR}$ and $r_{SLR}$ stand for the transmission and reflection coefficients of a single SLR, $\beta_2$ is the propagation constant of the multimode waveguide, $a_2$ represents the intracavity amplitude attenuation factor between the two mirrors; and $l_2$ is the cavity length separating the two SLRs. The normalized intensity response of the resonant system is given by：:

$$I_{\mathrm{out}} = \left|T_{\mathrm{FP}}\right|^2 \tag{4}$$

The SLR's reflectivity is evaluated through the combined FDTD and frequency domain simulations using a commercial software (Lumerical Inc.). We first simulate the behaviors of the 3-dB directional coupler for both $TM_0$ and $TE_0$ modes. The two parallel waveguides are 58.2 µm long and spaced by 350 nm, related to the beating length of the directional coupler. The S-parameters of the directional coupler, extracted from Lumerical's FDTD, are imported into INTERCONNECT for modelling the SLR's behaviors. As shown in Fig. 2, the $TM_0$ mode achieves near-unity reflectance at 1622 nm, while the $TE_0$ mode exhibits lower reflectance, leading to a reduced Q-factor in the corresponding resonator.

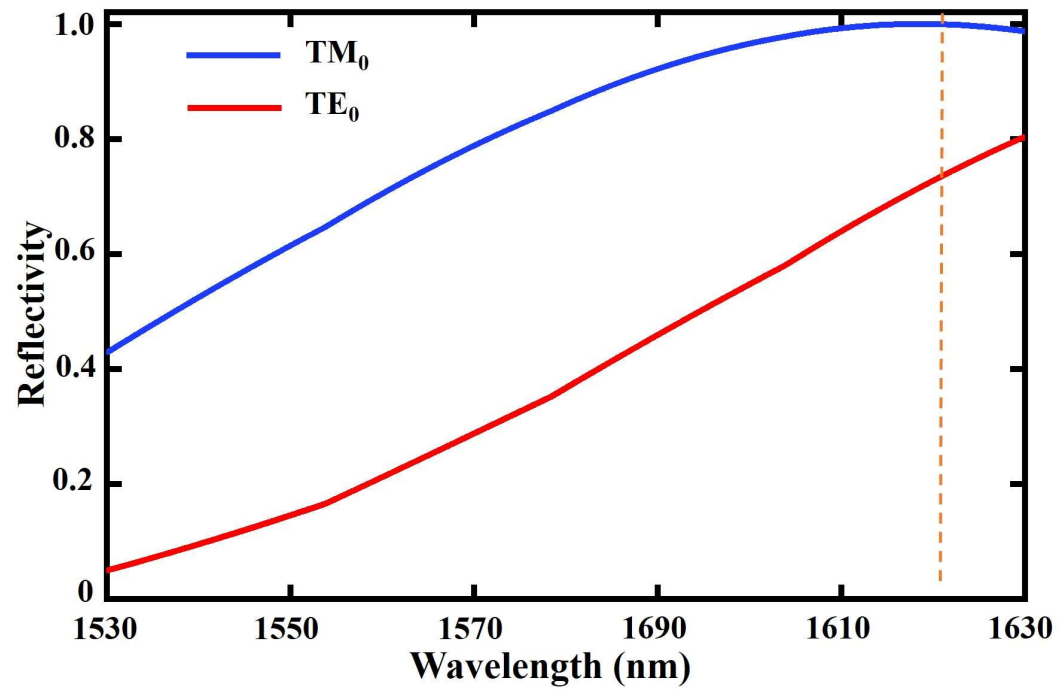

**Fig. 2** The simulated reflection spectra of the designed SLR for $TM_0$ and $TE_0$ modes for the wavelength spanning from 1530 nm to 1630 nm.

## 3. EXPERIMENT RESULTS

The schematic diagram of the experimental setup for characterizing the transmission characteristics of the µFP resonator is illustrated in Fig. 3. The light emitted from a tunable laser source (Santec TLS-570) is coupled into the chip via a lensed fiber. After passing through the device under test, the transmitted light is collected by an aspheric lens and directed into an optical polarizer (OPPF05-NIR, JCOPTIX). By rotating its polarization axis, the transmission spectra of TE and TM polarizations are measured separately, which are subsequently detected by a photodetector module (Santec MPM-210). After referencing the inverse tapers' coupling losses, all measured transmission spectra are normalized with respect to those obtained from a U-shaped reference waveguide as shown in the top inset of Fig.3.

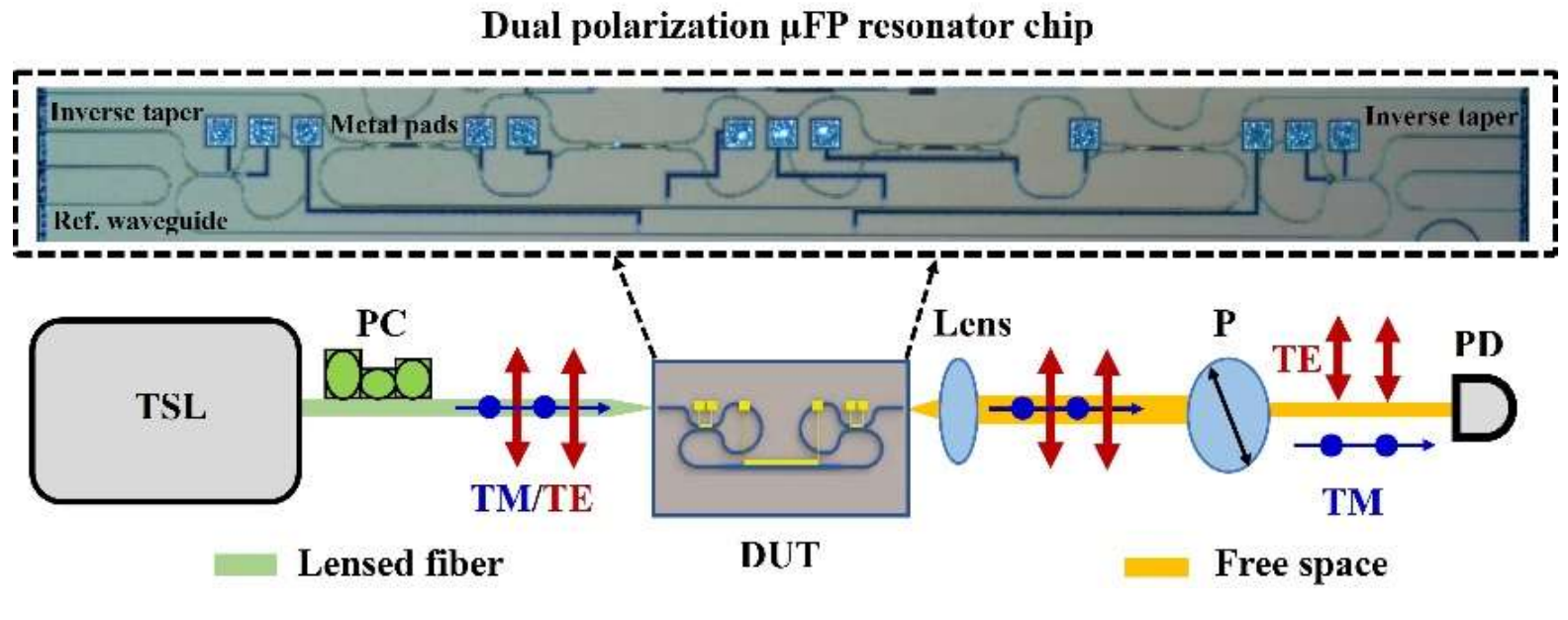

**Fig. 3** The schematic diagram of the experimental setup for device characterizations (TSL: tunable laser source, PC: polarization controller; DUT: device under test; P: polarizer; PD: photodetector).

We measured a loaded Q-factor exceeding $2.38 \times 10^6$ for the $TM_0$ mode of the fabricated device near 1625 nm. Fig. 4(a) presents the experimental transmission spectrum for the µFP resonator's $TM_0$ mode and the inset reveals an FSR of ca. 0.111 nm and an extinction ratio reaching up to 35 dB. The Lorentzian curve fitting was performed on the resonance peak centered at 1625.2486 nm, as illustrated in Fig. 4(b), yielding a full width at half maximum (FWHM) of 0.91 pm and a corresponding loaded Q-factor of around $2.38\times 10^6$. To the best of our knowledge, this represents the highest reported Q-factor for TM-polarized µFP resonators to date. To comprehensively evaluate the Q-factor variations, resonance peaks within the wavelength range from 1580 nm to 1630 nm were systematically analyzed. The fitting results, depicted in Fig. 4(c), indicate that more than 108 resonances feature at least a million Q-factors and the minimum Q-factor is approximately 2.0×105 near 1580 nm. As the wavelength approaches 1625 nm, the loaded Q-factors increase up to the peak. This wavelength-dependent behavior is attributed to the modulated SLR mirror reflectivity from 3-dB directional couplers. The bandwidth limitation of the prevailing 3-dB directional couplers can be significantly enhanced by implementing the curved waveguide configurations [31].

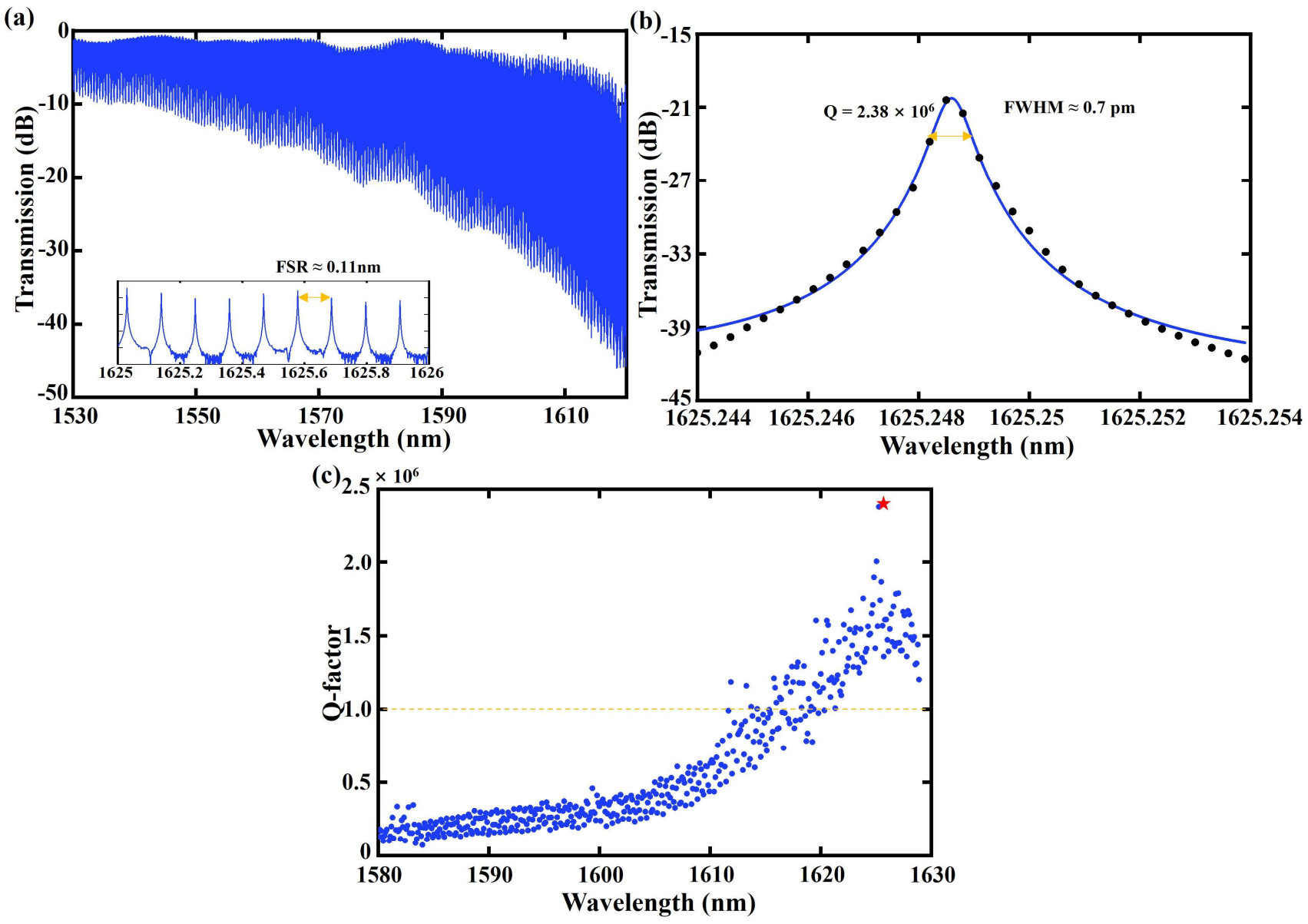

**Fig. 4** (a) Transmission spectrum of the μFP resonator of the $TM_0$ mode with an insertion loss of approximately 2 dB. The inset presents a magnified view of one selected spectral region with the indication of its FSR. (b) Lorentzian fitting of a single resonance peak with the highest Q-factor with the indication of its FWHM. (c) Q-factors of resonance peaks extracted from Lorentzian fits for the wavelength from 1580 nm to 1630 nm.

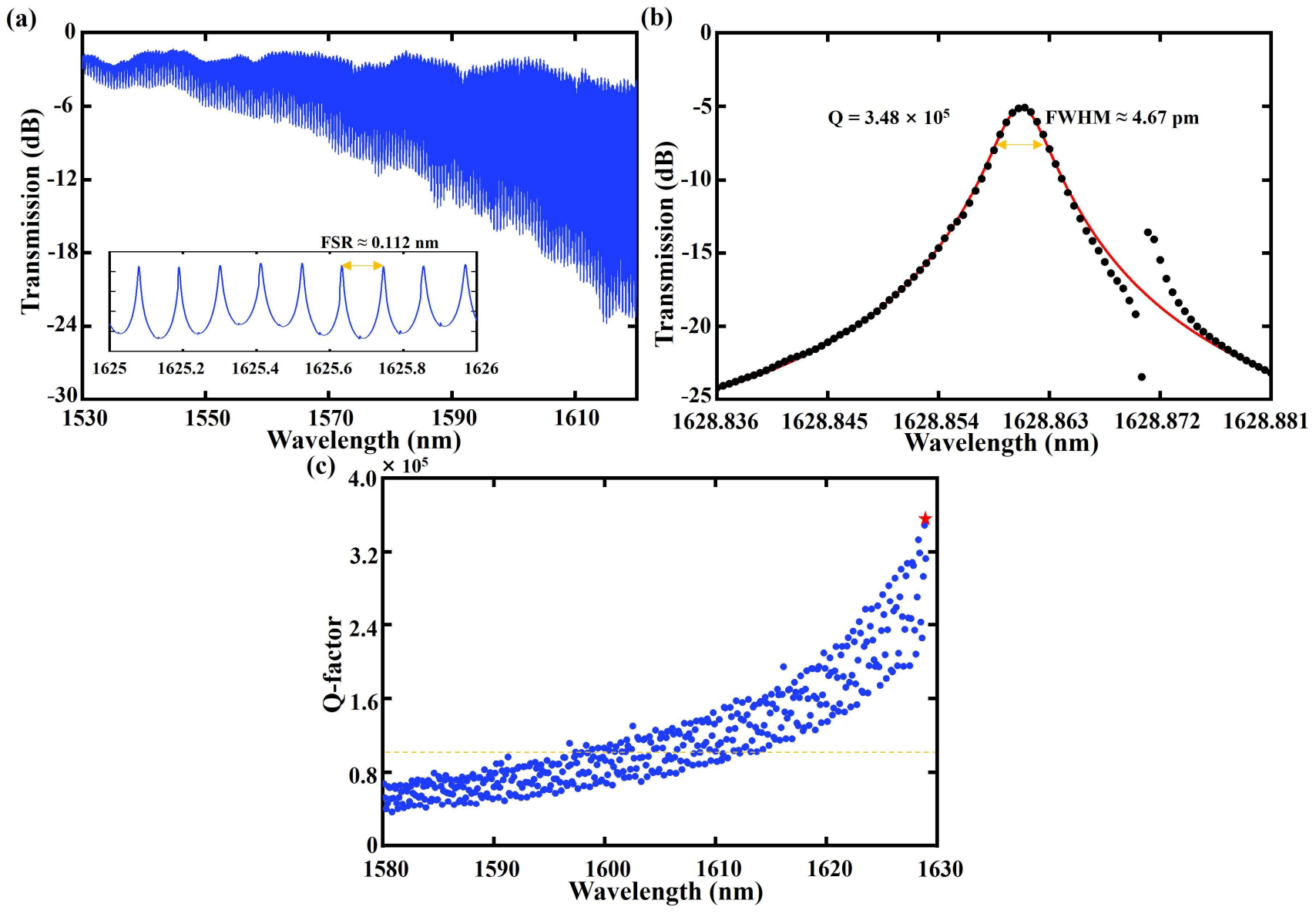

**Fig. 5** (a) Transmission spectrum of the μFP resonator of the $TE_0$ mode with an insertion loss of approximately 2.5 dB. The inset presents a magnified view of the selected spectral region. (b) Lorentzian fitting of a single resonance peak with the highest Q-factor with the indication of its FWHM. (c) Q-factors of resonance peaks extracted from Lorentzian fits for the wavelength from 1580 nm to 1630 nm.

To validate the polarization insensitive performance, we show that the loaded Q-factor for the $TE_0$ mode of the fabricated device can be as high as $3.48 \times 10^5$. Similarly, the measured transmission spectrum is shown in Fig. 5(a) after changing the polarization to the TE mode subsequently. The zoomed-in view shows a FSR of ca. 0.112 nm and an extinction ratio near 20 dB in the long-wavelength band. Fig. 5(b) displays the Lorentzian fitting of a single TE resonance peak at 1628.8607 nm, highlighting a fitted loaded Q-factor of 3.48 × $10^5$ and a FWHM of 4.67 pm. The Fano-like resonance at this resonance tail (1628.872 nm) arises due to the interplay with the suppressed $TM_0$ resonance [32]. In the same way, all the resonances from the 50-nm-wide spectral range were fitted, confirming that most of Q-factors exceeding 105, comparable to $TM_0$ mode performance. The Q-factor discrepancy between modes primarily stems from imbalanced power splitting in the 3-dB directional coupler, attributed to non-square waveguide geometry. Adopting the birefringence minimized directional couplers design reported in [33] may mitigate this issue.

We show that both the $TM_0$ and $TE_0$ modes are thermally tuned across one FSR while preserving their Q-factors. Heating was applied to the multimode waveguide region, modulating only the effective cavity length with negligible impact on the reflections at the two SLRs. As shown in Fig. 6(a), the resonance peak wavelength of the selected $TM_0$ mode shifted from 1607.665 nm to 1607.7483 nm, achieving a tuning range of roughly 0.083 nm at 80.4 mW power consumption. Similarly, Fig. 6(b) shows the $TE_0$ resonance tuning from 1607.6483 nm to 1607.7479 nm under the same power. The linear least-squares fitting of the resonance wavelength shift versus tuning power, shown in Fig. 6(c), yields tuning efficiencies of 1.04 pm/mW ($TM_0$) and 1.24 pm/mW ($TE_0$), with high coefficients of determination ($r^2$=0.9982). Furthermore, the power consumption required to achieve a 2π phase shift is below 125 mW for both polarizations. These results indicate that the photonic resonator offers highly linear and predictable thermo-optic tuning for both orthogonal polarization states, highlighting its potential for integrated photonic applications demanding precise wavelength control and spectral reconfigurability.

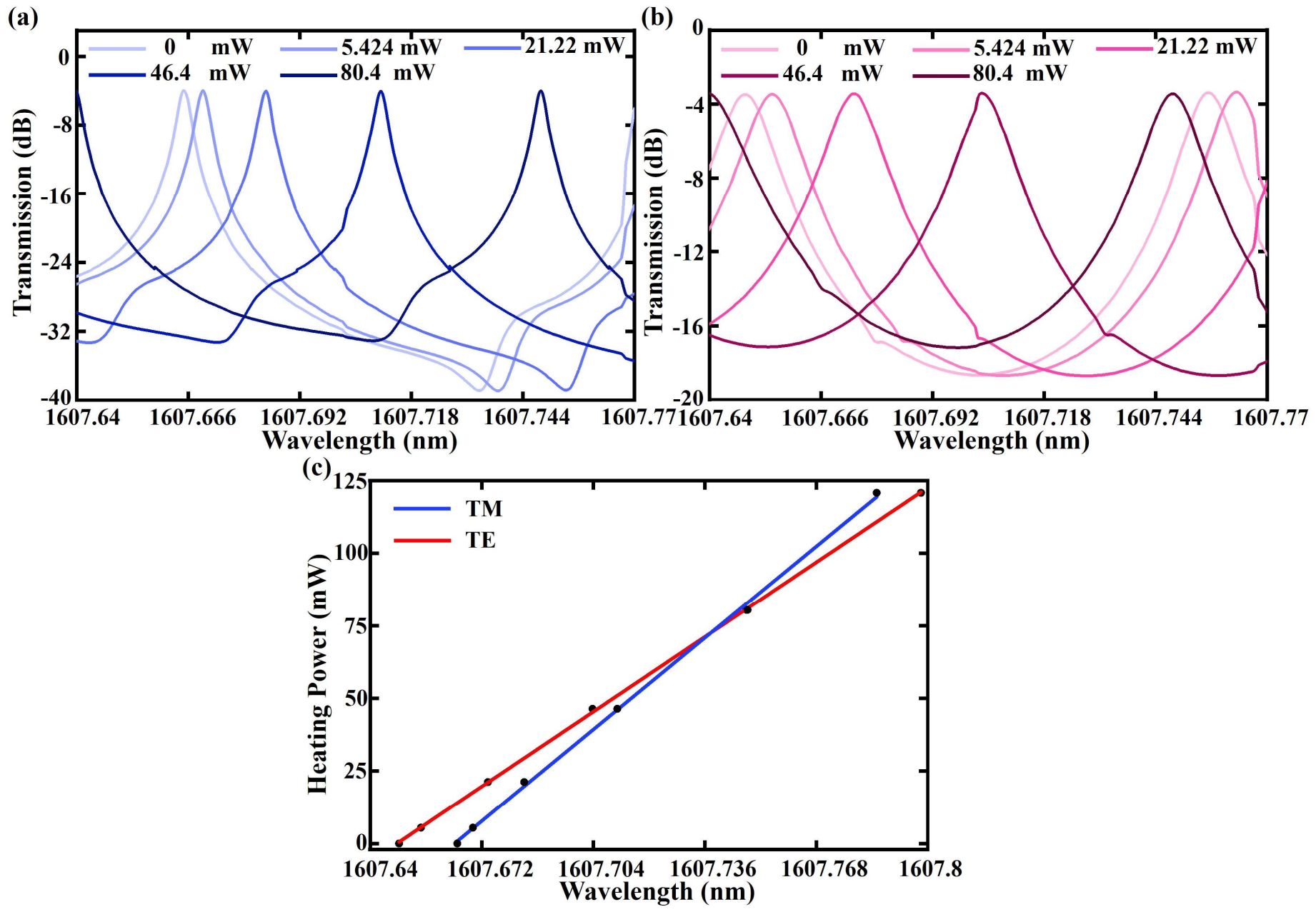

**Fig. 6** The transmission spectra under different electrical powers of (a) the $TM_0$ mode; (b) the $TE_0$ mode. (c) The resonance peak wavelength changes as a function of the electrical power for the $TM_0$ and $TE_0$ modes.

We compare the performance of various on-chip μFP and microring resonators (MRRs) operating at the near-infrared spectral range across diverse material platforms in Table 1. To the best of our knowledge, this work reports the first integrated dual-polarized μFP cavity achieving a record-high experimentally demonstrated Q-factor of 2.38×$10^6$ for the $TM_0$ mode. Further enhancement of the Q-factor is feasible through several strategies, including implementing adiabatic 3-dB power splitters to minimize amplitude imbalance, utilizing Euler-bend waveguides [34], [35] in the Sagnac loop to suppress bending loss, and optimizing the fabrication processes to reduce the sidewall roughness-induced scattering loss. Dual-polarization MRRs have been reported on SiC, $Si_3N_4$, and $SiO_2$ platforms [17] [48] [49], with record Q-factor up to 2×$10^8$ [49]. However, as traveling-wave resonators, conventional MRRs lack intrinsic reflection and require external feedback loops to provide deterministic frequency-selective feedback for self-injection locking [17]. This introduces additional loss (reducing effective Q-factors) and system complexity. In contrast, μFP resonators inherently generate selective feedback at every resonance mode due to their reflective boundaries, enabling direct high-stability self-injection locking without auxiliary components.

Table 1 PERFORMANCE COMPARISON OF ON-CHIP FP CAVITIES WORKING AT INFRARED WAVELENGTHS

| Reference | Reflector Type | FSR [nm] | Q-factor | Size | Platform |
|---|---|---|---|---|---|
| [8] | Retroreflector | 0.16 | 2.1×$10^6$ (TE) | 3.09 × 0.02 $mm^2$ | SOI |
| [36] | Inverse design | 0.16 | $2.03\times10^6$ (TE) | / | SOI |
| [37] | Inverse design | 10 | $2\times10^3$ (TE) | 2.16 × 2.16 $\mu m^2$ | SOI |
| [38] | SLR | 0.92 | $5.7\times10^4$ (TE) | / | SOI |
| [39] | Micro-ring assisted SLR | / | $2.8\times10^4$ (TE) | / | SOI |
| [40] | SLR | 0.28 | $6\times10^5$ (TE) | 1.06 × 0.28 $mm^2$ | SOI |
| [41] | DBR | / | $3.35\times10^4$ (TE) | 1.58 × 0.13 $mm^2$ | SOI |
| [42] | Photonic crystal Bragg grating reflector | 36 | $1.7\times10^3$ (TE) | 34.5×1$\mu m^2$ | SOI |
| [43] | Photonic crystal Bragg grating reflector | / | $1.72\times10^3$ (TE) | 6×1 $\mu m^2$ | SOI |
| [44] | DBR | / | $1.2\times10^4$ (TE) | 257 $\mu m^2$ | LNOI |
| [45] | SLR | 0.063 | $1.6\times10^5$ (TE) | 6.5 ×1.5 $mm^2$ | LNOI |
| [46] | SLR | / | $1.16\times10^6$ (TE) | / | LNOI |
| [47] | DBR | 1.39 | $4.8\times10^5$ (TE) | / | $Si_3N_4$ |
| [29] | DBR | / | $1.93\times10^7$ (TE) | / | $Si_3N_4$ |
| [9] | Undercut retroreflector | / | $5.4\times10^5$ (TE at 780 nm)<br>$1.4\times10^5$ (TE at 1310 nm) | / | $Si_3N_4$ |
| [48] | MRR | TM: ~3.3<br>TE: ~3.5 | $7.3\times10^4$ (TM)<br>$6.7\times10^4$ (TE) | 0.005 $mm^2$ | SiC |
| [17] | MRR | TM: 0.030<br>TE: 0.029 | $9\times10^6$ (TM)<br>$3.3\times10^7$ (TE) | 229 $mm^2$ | $Si_3N_4$ |
| [49] | MRR | TM: /<br>TE: ~0.117 | $2\times10^8$ (TM)<br>$1.2\times10^8$ (TE) | 14.54 $mm^2$ | $SiO_2$ |
| this work | Polarization-insensitive SLR | TM: 0.111<br>TE: 0.112 | $2.38\times10^6$ (TM)<br>$3.48\times10^5$ (TE) | 4.42 ×0.35 $mm^2$ | $Si_3N_4$ |

## 5. CONCLUSION

In summary, we demonstrate the design and realization of an ultra-high-Q dual-polarization μFP resonator based on a commercial $Si_3N_4$ platform. Its Q-factors of both $TM_0$ and $TE_0$ modes are significantly improved through two key strategies: polarization-insensitive SLRs are meticulously designed to minimize the reflection losses; the multimode waveguide is employed to reduce propagation losses. The proposed μFP resonator not only realizes simultaneous ultra-high-Q for two polarization modes but also achieves the thermo-optic tuning over one FSR. Experimentally, the $TM_0$ mode achieved a maximum loaded Q-factor of $2.38\times10^6$ with a FSR of 0.111 nm, while the $TE_0$ mode exhibited a Q-factor of $3.48\times10^5$ with a FSR of 0.112 nm. The intrinsic Q-factors for both $TM_0$ and $TE_0$ modes are even higher as commonly known. Furthermore, we demonstrate the gapless thermal tuning with the efficiencies of 1.04 pm/mW ($TM_0$) and 1.24 pm/mW ($TE_0$), respectively. These findings highlight its promising potential for applications in sensing, nonlinear and quantum photonics.

## 6. DISCUSSION

Although the measured loaded Q-factor ($Q_l$) of the dual-polarization μFP resonator has reached an exceptionally high level, its intrinsic Q-factor ($Q_i$) is higher and more accurately reflects the device's inherent loss characteristics. Due to MPW's space limitation, we do not have design variations to extract the resonator's $Q_i$. Currently, there are two approaches to determine $Q_i$. The first one is systematic measurements of resonators with multiple cavity lengths, combined with theoretical modeling to fit the dependence of $Q_l$ on the cavity length [50]. This approach enables the effective separation of propagation loss (α) and coupling loss (γ), thereby allowing the estimation of the $Q_i$ unaffected by coupling losses. The second one is the transfer matrix method based on the ABCD formalism [29] for $Q_i$ extraction. It enables precise numerical simulations and fitting of the transmission spectra, which restore the resonance peak lineshape and assist in distinguishing different loss contributions. The combined application of these two methods facilitates a comprehensive analysis of the intrinsic Q-factor. Future work could integrate multimodal measurement techniques and simulation methods to systematically investigate the factors influencing the $Q_i$, thereby advancing the ultra-high Q μFP cavity development.

### Declaration of Competing Interest

The authors declare that they have no known competing financial interests or personal relationships that could have appeared to influence the work reported in this paper.

### Data availability

Data will be made available on request.

### Acknowledgment

This work was supported by the National Natural Science Foundation of China under Grant 62105061 and 12374301, in part by the National Key Research and Development Program of China 2024YFA1210500 and by the Key Lab of Modern Optical Technologies of Education Ministry of China, Soochow University.